\documentclass[a4paper]{article}
\usepackage{graphicx}
\usepackage{amsmath}
\title{Basic logic and quantum entanglement}

\author{P A Zizzi}

\date{\ }

\begin{document}

\maketitle

\vspace{-1cm}

\begin{center}
\emph{Dipartimento di Matematica Pura ed Applicata\\
Via Trieste, 63 - 35121 Padova, Italy\\
e-mail: zizzi@math.unipd.it}
\end{center}

\vspace{0.5cm}

{\abstract

As it is well known, quantum entanglement is one of the most
important features of quantum computing, as it leads to massive
quantum parallelism, hence to exponential computational speed-up.
In a sense, quantum entanglement is considered as an implicit
property of quantum computation itself. But... can it be made
explicit? In other words, is it possible to find the connective
``entanglement'' in a logical sequent calculus for the machine
language? And also, is it possible to ``teach'' the quantum
computer to ``mimic'' the EPR ``paradox''? The answer is in the
affirmative, if the logical sequent calculus is that of the
weakest possible logic, namely Basic logic.\\ A weak logic has few
structural rules. But in logic, a weak structure leaves more room
for connectives (for example the connective ``entanglement'').
Furthermore, the absence in Basic logic of the two structural
rules of contraction and weakening corresponds to the validity of
the no-cloning and no-erase theorems, respectively, in quantum
computing. }

\section{Introduction}

Our purpose is to obtain an adequate sequent calculus \cite{Ge69}
for quantum computation \cite{NiChu00}. In particular, we look for
logical connectives corresponding to the physical links existing
among qubits in the quantum computer, and the associated inference
rules. To this aim, we will exploit Basic logic \cite{SBF00} and
its reflection principle between meta-language and object
language. The sequent calculus we are looking for should be able
to reproduce two main features of quantum computing namely quantum
superposition and quantum entanglement. These two features taken
together (the so-called quantum massive parallelism) are in fact
very important as they lead to quantum computational speed-up
\cite{JoLi??}. A logical interpretation of quantum superposition
is straightforward in Basic logic, and is given in terms of the
additive connective \&=``with'' (and of its symmetric,
$\vee$=``or'') both present in linear \cite{Gi87} and Basic
logics.

In this paper, we also propose a logical interpretation for
quantum entanglement. Entanglement is a strong quantum
correlation, which has no classical analogous. Then, the logic
having room for the connective ``entanglement'', will be selected
as the most adequate logic for quantum mechanics, and in
particular for quantum computing. Quantum entanglement is
mathematically expressed by particular superposition of tensor
products of basis states of two (or more) Hilbert spaces such that
the resulting state is non-separable. For this reason, one can
expect that the new logical connective, which should describe
entanglement, will be both additive and multiplicative, and this
is in fact the case. We introduce the connective
@=``entanglement'' by solving its definitional equation, and we
get the logical rules for @. It turns out that @ is a (right)
connective given in terms of the (right) additive conjunction \&
and of the (right) multiplicative disjunction $\wp$=``par''.

Then, we discuss the properties of @. In particular, we prove that
@ is not idempotent, which is equivalent to formulate the ``no
self-referentiality'' theorem in the meta-language.

Also, we show that, like all the connectives of Basic logic, @ has
its symmetric, the (left) connective \S,  given in terms of the
(left) additive disjunction, $\vee$=``or'', and  the (left)
multiplicative conjunction $\otimes$=``times'' (the symmetric of
$\wp$).

Moreover, we provide Basic logic of a new meta-rule, which we name
EPR-rule as it is the logical counterpart of the so-called EPR
``paradox'' \cite{EPR35}.

The conclusion of this paper is that Basic logic is the unique
adequate logic for quantum computing, once the connective
entanglement and the EPR-rule are included.

\section{A brief review of basic logic}

Basic logic \cite{SBF00} is the weakest possible logic (no
structure, no free contexts) and was originally conceived
\cite{BaSa97} as the common platform for all other logics
(linerar, intuitionistic, quantum, classical etc.) which can be
considered as its ``extensions''. Basic logic has tree main
properties:

\begin{itemize}

\item[\textbf{i)}]
\textbf{Reflection}: All the connectives of Basic logic satisfy
the principle of reflection, that is, they are introduced by
solving an equation (called \emph{definitional equation}),  which
``reflects'' meta-linguistic links between assertions into the
object-language.  There are only two metalinguistic links:
``yields'', ``and''. The metalinguistic ``and'', when is outside
the sequent, is indicated by \underline{and}; when inside the
sequent, is indicated by a comma.

$$
\begin{array}{ccl}
\scriptstyle
\textbf{\footnotesize{Object language}} & \underset{\scriptscriptstyle Reflection}{\longleftrightarrow}(=\mathit{iff}) & \textbf{\footnotesize{Meta-language}}\\
\text{\footnotesize{connectives}} & &
\text{\footnotesize{meta-linguistic links}}
\begin{cases}
\scriptstyle yelds & \vdash\\
\scriptstyle ``and\text{''} & \scriptstyle
\begin{cases}
\scriptstyle
\footnotesize{\text{\underline{and}}} & \scriptstyle outside \,\, the \,\, sequent\\
\scriptstyle , (comma) & \scriptstyle inside \,\, the \,\, sequent
\end{cases}
\end{cases}
\end{array}
$$

\item[\textbf{ii)}]
\textbf{Symmetry}: All the connectives are divided into ``left''
and ``right'' connectives.

A left connective has formation rule acting on the left, and a
reflection rule acting on the right. In Basic logic, every left
connective has its symmetric, a right connective, which has a
formation rule acting on the right, and a reflection rule acting
on the left (and vice-versa).

$$
\begin{array}{lcl}
\textbf{\footnotesize{Left connectives}} & \underset{\scriptscriptstyle Symmetry}{\longleftrightarrow} & \textbf{\footnotesize{Right connectives}}\\
\scriptstyle \vee=``or\text{''} \text{\footnotesize{(Additive
disjunction)}} & & \scriptstyle \&=``with\text{''}
\text{\footnotesize{(Additive conjunction)}}\\
\scriptstyle \otimes=``times\text{''}
\text{\footnotesize{(Multiplicative conjunction)}} & &
\scriptstyle \wp=``par\text{''}
\text{\footnotesize{(Multiplicative disjunction)}}\\
\scriptstyle \leftarrow \,
\text{\footnotesize{(Counterimplication)}} & & \scriptstyle
\rightarrow \,
\text{\footnotesize{(Implication)}}\\
\end{array}
$$

\item[\textbf{iii)}]
\textbf{Visibility}: There is a strict control on the contexts,
that is, all active formulas are isolated from the contexts, and
they are \emph{visible}.

In Basic logic, the identity axiom $A \vdash A$ and the cut rule:
$\displaystyle \frac{\Gamma \vdash A \;\; A \vdash \Delta}{\Gamma
\vdash \Delta}\, cut$ hold.

The cube of logics \cite{SBF00} \cite{BaSa97} is a geometrical
symmetry in the space of logics, which becomes apparent once one
takes Basic logic as the fundamental one. As we said above, Basic
logic is the weakest logic, and all the other logics can be
considered as its extensions. All the logics, which have no
structural rules (called substructural logics or resources logics)
are the four vertices of one same face of the cube, considered as
the basis. They have many connectives, less structure, and less
degree of abstraction. On the upper face of the cube, we have all
the structural logics (they have fewer connectives, more
structure, and a higher degree of abstraction). See Fig.~1.

\begin{figure}[h]
\label{fig1}
\includegraphics[width=18pc]{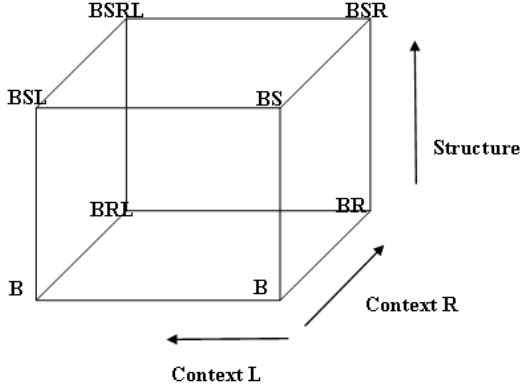}
\caption{The cube of logics}
\end{figure}

\begin{tabular}{ll}
\textbf{Substructural logics:} & \textbf{Structural logics:}\\
\textbf{B} = Basic logic & \textbf{BS} = quantum logic\\
\textbf{BL} = Basic logic + context on the left & \textbf{BSR} = paraconsistent logic\\
\textbf{BR} = Basic logic + context on the right & \textbf{BSL} =
intuitionistic logic\\
\textbf{BRL} = Linear logic & \textbf{BSRL} = classical logic\\
\end{tabular}

\end{itemize}

\section{Reasons why Basic logic is the logic of quantum
computing} \label{sec3}

Basic logic has the following features, which are essential to
describe quantum computation in logical terms:

\begin{itemize}

\item[a)]  It is \textbf{non-distributive} (because of the absence,
on both sides of the sequent, of active contexts), and this is of
course a first necessary requirement for any logic aimed to
describe a quantum mechanical system.

\item[b)]  It is \textbf{substructural}, i.e., it has no structural
rules like contraction: $\displaystyle
\frac{\Gamma,A,A\vdash\Delta}{\Gamma,A\vdash\Delta}$ (data can be
copied) and weakening: $\displaystyle
\frac{\Gamma\vdash\Delta}{\Gamma,A\vdash\Delta}$  (data can be
deleted), accordingly with the no-cloning  theorem \cite{WoZu82}
and the no-erase theorem \cite{PaBra00} respectively, in quantum
computing. The only structural rule, which holds in Basic logic,
is the exchange rule:

$$
exchL\,
\frac{\Gamma,A,B,\Gamma'\vdash\Delta}{\Gamma,B,A,\Gamma'\vdash\Delta}
\qquad\qquad exchR\,
\frac{\Gamma\vdash\Delta,A,B,\Delta'}{\Gamma\vdash\Delta,B,A,\Delta'}
$$

Then, for example, standard quantum logic \cite{BiNe36} although
being non-distributive, is excluded as a possible candidate for
quantum computing because it has structural rules.

Linear logic is substructural, but has both left-side and
right-side free contexts, then is excluded because of
distributivity. (In particular, as we will see, for the connective
@ =``entanglement'', the distributive property does not hold, then
Linear logic cannot accommodate @).

\item[c)] It is \textbf{paraconsistent}: the non-contradiction
principle is invalidated, and quantum superposition can be assumed.

\end{itemize}

\section{The logical connective \& for quantum superposition}

The unit of quantum information is the qubit
$|Q\rangle=a|0\rangle+b|1\rangle$, which is a linear combination
of the basis states $|0\rangle$ and $|1\rangle$, with complex
coefficients $a$ and $b$ called probability amplitudes, such that
the probabilities sum up to one: $|a|^2+|b|^2=1$. In logical
terms, we will interpret the atomic proposition $A$ as bit
$|1\rangle$, and its primitive negation $A^{\perp}$ as bit
$|0\rangle=NOT|1\rangle$, where $NOT$ is the $2\times 2$
off-diagonal matrix $\displaystyle NOT=
\begin{pmatrix} 0 & 1\\ 1 & 0
\end{pmatrix}$.

The atomic assertion $\vdash A$  will be interpreted as the
quantum state $\vdash A\equiv |A\rangle=b|1\rangle$,  and the
asserted negation as $\vdash (A^{\perp})\equiv |A^{\perp}\rangle=a
NOT|1\rangle=a|0\rangle$. Notice that making the negation of the
atomic assertion $\vdash A$ is not the same as asserting the
negation of the atomic proposition, in fact $(\vdash A)^\perp
\equiv NOT|A\rangle=b|0\rangle$.

In the meta-language, quantum superposition means that both
propositions $A$ and $A^\perp$ are asserted, that is, on the
right-hand side of the definitional equation, we will have:
$\vdash A$ \underline{and} $\vdash (A^\perp)$. On the left-hand
side, we look for the connective \$=``superposition'', such that
$\vdash A \$ A^\perp$ \underline{iff} $\vdash A$ \underline{and}
$\vdash (A^\perp)$.

This is the definitional equation of \& \cite{SBF00}:

$$
\Gamma \vdash A \& B \qquad \mathrm{\underline{iff}} \qquad
\Gamma\vdash A \quad \mathrm{\underline{and}} \quad \Gamma\vdash B
$$

in the particular case with $B=A^\perp$ and $\Gamma=\emptyset$.

So that we can write the definitional equation for the connective
``quantum superposition'' as:

\begin{equation}
\label{eq1} \vdash A\& A^\perp \qquad \mathrm{\underline{iff}}
\qquad \vdash A \quad \mathrm{\underline{and}} \quad
\vdash(A^\perp)
\end{equation}

Then, of course, the rules of the connective ``quantum
superposition'' are the same rules of \& \cite{SBF00}, with
$B=A^\perp$ and $\Gamma=\emptyset$.

$$
\&- \, form \, \frac{\vdash A\quad\vdash(A^\perp)}{\vdash A\& B}
$$

This is obtained from the RHS to the LHS of the definitional
equation \eqref{eq1}.

$$
\& - \, implicit \; refl \, \frac{\vdash A\& A^\perp}{\vdash A}
\quad \frac{\vdash A\& A^\perp}{\vdash (A^\perp)}
$$

This is obtained from the LHS to the RHS of the definitional
equation \eqref{eq1}.

By trivializing the \&-implicit reflection, i.e., putting
$\Gamma=A \& A^\perp$, we get the two \&-axioms:

$$
A\& A^\perp \vdash A, \qquad A\& A^\perp \vdash A^\perp
$$

Suppose now $A\vdash \Delta (A^\perp\vdash\Delta)$. By composition
with the ``axiom'' $A\& A^\perp\vdash A (A\& A^\perp \vdash
A^\perp)$ we get the \&-explicit reflection rule:

$$
\& - \,\mathrm{exp}l.refl\, \frac{A\vdash \Delta}{A\& A^\perp
\vdash \Delta} \qquad
\frac{A^\perp\vdash\Delta}{A\&A^\perp\vdash\Delta}
$$

As we have completely solved the definitional equation \eqref{eq1}
we can express quantum superposition in the object language with
the composite proposition $A\&A^\perp$. Asserting: $A\&A^\perp$
(i.e. $\vdash A\& A^\perp$) is then equivalent to $(\vdash
A)\&(\vdash A^\perp)$.

The logical expression of the qubit
$|Q\rangle=a|0\rangle+b|1\rangle$ is then:

\begin{equation}
\label{eq2} \vdash Q\equiv\vdash A\&A^\perp
\end{equation}

\section{The logical connective @ for quantum entanglement}

Two qubits $|Q\rangle_A=a|0\rangle_A+b|1\rangle_A$,
$|Q\rangle_B=a'|0\rangle_B+b'|1\rangle_B$ are said entangled when
the two qubits state $|Q\rangle_{AB}$  is not separable, i.e.,
$|Q\rangle_{AB}\neq |Q\rangle_A\otimes |Q\rangle_B$, where
$\otimes$ is the tensor product in Hilbert spaces. In particular,
a two qubit state is maximally entangled when it is one of the
four Bell states \cite{Be87}:

$$
|\Phi_\pm\rangle_{AB} = \frac{1}{\sqrt{2}}\left(|0\rangle_A
\otimes |0\rangle_B \pm |1\rangle_A\otimes|1\rangle_B\right),\quad
|\Psi_\pm\rangle_{AB}=\frac{1}{\sqrt{2}}
\left(|0\rangle_A\otimes|1\rangle_B\pm|1\rangle_A\otimes|0\rangle_B\right).
$$

For simplicity, in this paper we will consider only Bell states.
As we have seen in Sect.~\ref{sec3}, expressing the qubit
$|Q\rangle_A$ in logical terms leads to the compound proposition
$Q_A\doteq A\& A^\perp$, where \& stands for the connective
``and''. In the same way, we can associate a proposition $B$ to
the bit $|1\rangle_B$ and its primitive negation $B^\perp$ to the
bit $|0\rangle_B$  so that the second qubit $|Q\rangle_B$ is
expressed, in logical terms, by a second compound proposition
$Q_B\doteq B\& B^\perp$. Bell states will be expressed, in logical
terms, by the expression $Q_A@Q_B$, where @  is the new logical
connective called ``entanglement''. Like all the other
connectives, @ will be defined by the reflection principle, which
translates meta-language into object language. We have at our
disposal a meta-language which comes from our knowledge of the
physical structure of Bell states. This leads us to figure out the
logical structure for, say, the Bell states
$|\Phi_\pm\rangle_{AB}$, namely $\vdash(A\wp B)\&(A^\perp\wp
B^\perp)$.  Similarly, the logical structure for the Bell states
$|\Psi_\pm\rangle_{AB}$ will be: $\vdash(A\wp
B^\perp)\&(A^\perp\wp B)$. In the following, we will consider only
the logical expression for the states $|\Phi_\pm\rangle_{AB}$, as
the case for $|\Psi_\pm\rangle_{AB}$ is obtained exchanging $A$
with $A^\perp$. Eventually, we get the following definition: Two
compound propositions $Q_A\doteq A\& A^\perp$, $Q_B\doteq B\&
B^\perp$ will be said (maximally) entangled if they are linked by
the connective @ = ``entanglement''. The definitional equation for
@ is:

\begin{equation}
\label{eq3} \Gamma \vdash Q_A@Q_B \qquad \mathrm{\underline{iff}}
\qquad \Gamma \vdash A,B \quad \mathrm{\underline{and}} \quad
\Gamma\vdash A^\perp,B^\perp
\end{equation}

On the right-hand side of the definitional equation, we have the
meta-language, coming from our knowledge of the physical structure
of Bell states. On the left-hand side, instead, we have the object
language. Also, it should be noticed that, on the right hand side
of the definitional equation, each of the two commas is reflected
into a $\wp$ while the meta-linguistic link \underline{and} is
reflected into \& Thus the connective @ is an additive as well as
multiplicative connective (more exactly, an additive conjunction
and a multiplicative disjunction) which reflects two kinds of
``and'' on the right: one outside the sequent (\underline{and})
and one inside the sequent (the comma). Finally, the connective @,
is a derived connective  which, nevertheless, has its own
definitional equation: this is a new result in logic. Solving the
definitional equation for @ leads to the following rules:

\begin{align}
\label{eq4} @&-\text{formation} && \frac{\Gamma\vdash A,B \qquad
\Gamma\vdash A^\perp,B^\perp}{\Gamma\vdash Q_A@Q_B}&&\\
\label{eq5} @&-\text{implicit reflection} & &\frac{\Gamma\vdash
Q_A@Q_B}{\Gamma\vdash A,B}\;\text{(i)} & &\frac{\Gamma\vdash Q_A@
Q_B}{\Gamma\vdash A^\perp,B^\perp}\;\text{(ii)}\\
\label{eq6}@&-\text{axioms} & &Q_A@Q_B\vdash A,B\;\text{(i)} &&
Q_A@ Q_B\vdash A^\perp,B^\perp\;\text{(ii)}\\
\label{eq7}@&-\text{explicit reflection} &&
\frac{A\vdash\Delta\qquad B\vdash
\Delta'}{Q_A@Q_B\vdash\Delta,\Delta'}\;\text{(i)} &&
\frac{A^\perp\vdash\Delta \qquad
B^\perp\vdash\Delta'}{Q_A@Q_B\vdash\Delta,\Delta'}\;\text{(ii)}
\end{align}

Eq.~\eqref{eq4} is equivalent to the @-definitional equation from
the right hand side to the left hand side. Eqs.~\eqref{eq5} are
equivalent to the @-definitional equation from the left hand side
to the right hand side. The @-axioms  in \eqref{eq6} are obtained
from \eqref{eq5}, by the trivialization procedure, that is,
setting $\Gamma=Q_A@Q_B$. The @-explicit reflection rules (i),
(ii)  in \eqref{eq7} are obtained by composition of the @-axioms
(i) and (ii) in \eqref{eq6} with the premises $A\vdash\Delta$ and
$B\vdash\Delta'$, and $A^\perp\vdash\Delta$ and
$B^\perp\vdash\Delta'$, respectively.

The properties of @ are:

\begin{itemize}

\item[1)] \textbf{Commutativity:}
\begin{equation}
\label{eq8} Q_A@Q_B \doteq Q_B@Q_A
\end{equation}

Commutativity of @ holds if and only if, the exchange rule is
assumed (on the right). And in fact, exchange is a valid rule in
Basic logic.

\item[2)] \textbf{Semi-distributivity}

>From the definitional equation of @ with $\Gamma=\emptyset$, that
is:

$$
\vdash Q_A@Q_B \qquad \text{\underline{iff}} \qquad \vdash A,B
\quad \text{\underline{and}} \quad \vdash A^\perp,B^\perp
$$

we get:

\begin{equation}
\label{eq9} (A\&A^\perp)@(B\&B^\perp)\doteq (A\wp B)\&
(A^{\perp}\wp B^\perp)
\end{equation}

We see that two terms are missing in \eqref{eq9} namely $(A\wp
B^\perp)$ and $(A^\perp\wp B)$, so that @ has distributivity with
absorption, which we call semi-distributivity.

\item[3)] \textbf{Duality}

Let us define now the dual of @: $(Q_A@Q_B)^\perp\equiv [(A\wp
B)\&(A^\perp\wp B^\perp)]^\perp = (A\otimes B) \vee
(A^\perp\otimes B^\perp)$ and let us call it \S, that is:
$(Q_A@Q_B)^\perp\equiv Q_A\S Q_B$ (vice-versa, the dual of $\S$ is
@: $(Q_A\S Q_B)^\perp\equiv Q_A@Q_B)$.

The definition of the dual of @ is then:

\begin{equation}
\label{eq10} Q_A\S Q_B\doteq (A\otimes B)\vee(A^\perp\otimes
B^\perp)
\end{equation}

\item[4)] \textbf{Non Associativity:}

\begin{equation}
\label{eq11} Q_A@(Q_B@Q_C)\neq(Q_A@Q_B)@Q_C
\end{equation}

To discuss associativity of @, a third qubit $Q_C$  is needed, and
$Q_A@(Q_B@Q_C)\doteq(Q_A@Q_B)@Q_C$ cannot be demonstrated in Basic
logic, as $Q_C$ acts like a context on the right.

We remind that the maximally entangled state of three qubits is
the GHZ state \cite{GrHoShZe90}.

\item[5)] \textbf{Non-idempotence:}

\begin{equation}
\label{eq12} Q_A@Q_A\neq Q_A
\end{equation}

The proof of \eqref{eq12} and its interpretation will be given in
a forthcoming paper \cite{Zi??}.

\end{itemize}

\section{The EPR rule}

Let us consider the cut:

\begin{equation}
\label{eq13} \frac{\vdash Q_A \qquad Q_A\vdash A}{\vdash A}\, cut
\end{equation}

which corresponds, in physical terms, to measure the qubit
$|Q_A\rangle_A$ in state $|1\rangle_A$ (with probability $|b|^2$).
In the same way, the cut: $\displaystyle\frac{\vdash Q_A\quad
Q_A\vdash A^\perp}{\vdash A^\perp}\,cut$,  corresponds to measure
the qubit $|Q_A\rangle_A$ in state $|0\rangle_A$ (with probability
$|a|^2$).

The cut (over entanglement) is:

\begin{equation}
\label{eq14} \frac{\vdash Q_A@Q_B \qquad Q_A@Q_B\vdash A,
B}{\displaystyle \frac{\vdash A,B}{\vdash A\wp
B}\,\wp-formation}\,cut
\end{equation}

Where, in \eqref{eq14}, the rule of $\wp-fomation$ \cite{SBF00}
is: $\displaystyle\wp-form\,\frac{\Gamma\vdash A,B}{\Gamma\vdash
A\wp B}$.

Performing the cut in \eqref{eq14} corresponds, in physical terms,
to measure the state $|1\rangle_A|1\rangle_B$. If we replace $A$
and $B$ in \eqref{eq14} with $A^\perp$, $B^\perp$ the cut
corresponds to measure the state $|0\rangle_A |0\rangle_B$.  It
should be noticed, that, if we make a measurement of $Q_A$
(supposed entangled with $Q_B$) and get $A$, then by
semi-distributivity of @, we have:

\begin{equation}
\label{eq15} A@Q_B\equiv A@(B\&B^\perp)\doteq A\wp B
\end{equation}

As it is well known, if two quantum systems $S_A$ and $S_B$  are
entangled, they share a unique quantum state, and even if they are
far apart, a measurement performed on $S_A$ influences any
subsequent measurement performed on $S_B$  (the EPR ``paradox''
\cite{EPR35}). Let us consider Alice, who is an observer for
system $S_A$, which is the qubit $Q_A$, that is, she can perform a
measurement of $Q_A$. There are two possible outcomes, with equal
probability $1/2$:

\begin{itemize}
\item[(i)]
Alice measures 1, and the Bell state collapses to
$|1\rangle_A|1\rangle_B$.

\item[(ii)]
Alice measures 0, and the Bell state collapses to
$|0\rangle_A|0\rangle_B$.
\end{itemize}

Now, let us suppose Bob is an observer for system $S_B$ (the qubit
$Q_B$). If Alice has measured $1$, any subsequent measurement of
$Q_B$ performed by Bob always returns $1$. If Alice measured $0$,
instead, any subsequent measurement of $Q_B$ performed by Bob
always returns $0$. To discuss the EPR paradox in logical terms,
we introduce the EPR rule:

\begin{equation}
\label{eq16} \frac{\Gamma\vdash Q_A@Q_B \qquad Q_A\vdash
A}{\displaystyle \frac{\Gamma\vdash
A@Q_B}{\displaystyle\frac{\Gamma\vdash A,B}{\Gamma \vdash A\wp
B}\,\wp-form.}\,@-impl.refl.}
\end{equation}

Where the semi-distributivity of @, i.e. $A@Q_B\doteq A,B$  has
been used in the step ``@-impl.refl''.

Notice that the consequences of the EPR rule are the same of the
cut over entanglement \eqref{eq14}, because of semi-distributivity
of @. It was believed that no other rule existed, a part from the
cut rule, or at least some rule equivalent to it, which could cut
a formula in a logical derivation. Nevertheless, the EPR rule does
cut a formula, but it can be proved that it is not equivalent to
the cut rule over entanglement (and, vice-versa, the cut rule over
entanglement is not equivalent to the EPR rule). This is a new
result in logic.

Let us show first that the EPR rule is not equivalent to the cut
rule. We start with the premises of the EPR rule and apply the cut
rule:

\begin{equation}
\label{eq17} weak.L\, \frac{\displaystyle\frac{\Gamma\vdash
Q_A@Q_B}{\displaystyle \frac{\Gamma\vdash Q_A@Q_B \quad
Q_A@Q_B\vdash A,B}{\Gamma\vdash A,B}\,cut}\,@-axiom \qquad
Q_A\vdash A}{\displaystyle\frac{\Gamma\vdash A,B,Q_A \qquad
Q_A\vdash A}{\displaystyle\frac{\Gamma\vdash A,A,B}{\Gamma\vdash
A,B}\, contr.R}\,cut}
\end{equation}

It is clear that it is impossible to demonstrate that the EPR rule
is equivalent to the cut rule (over entanglement) in Basic logic,
where we don't have the structural rules of weakening and
contraction. And in any logic with structural rules, the
connective entanglement disappears, and the EPR rule collapses to
the cut rule.

Now, we will show the vice-versa, i.e., that the cut rule (over
entanglement) is not equivalent to the EPR rule. We start with the
premises of the cut rule \eqref{eq14} and apply the EPR rule
\eqref{eq16}:

\begin{equation}
\label{eq18} \frac{\Gamma \vdash Q_A@Q_B \qquad \displaystyle
\frac{Q_A@Q_B\vdash A,B}{Q_A@Q_B,Q_A\vdash A,B}\,weak.L}{\Gamma,
Q_A@Q_B\vdash A@Q_B,B}\,EPR^C
\end{equation}

Where in \eqref{eq18}, $EPR^C$ means EPR rule in presence of
contexts (here $Q_A@Q_B$  on the left and $B$ on the right). But
contexts are absent in Basic logic (visibility). Furthermore, the
weakening rule is not present in Basic logic. These facts lead to
the conclusion that in Basic logic it is impossible to prove that
the cut is equivalent to the EPR rule. Moreover, this is
impossible also in sub-structural rules with contexts (like BL,
BR, and BLR) because one cannot use weakening, and in structural
logics because the connective entanglement disappears.

The EPR rule is a new kind of meta-rule peculiar of entanglement,
which is possible only in Basic logic.  It is a stronger rule
(although less general) than the cut, as it uses a weaker premise
to yield the same result. Hence, instead of proving $Q_A@Q_B$ in
\eqref{eq14} that is $\vdash Q_A@Q_B$, we can just prove $Q_A$,
i.e., $\vdash Q_A$, perform the usual cut \eqref{eq13} (over
$Q_A$), and leave the result $A$ entangled with $Q_B$. Roughly
speaking, if two compound propositions are (maximally) entangled,
it is sufficient to prove only one of them. This is the logical
analogue of the EPR ``paradox''.

\section{Conclusions}

Basic logic, once endowed with the new connective ``entanglement''
and the EPR rule, provides the unique adequate sequent calculus
for quantum computing. We list below the main features of quantum
information and quantum computing, and the corresponding required
properties for the associated logic.
The main features of quantum
computing are:

\begin{itemize}
\item{1)} Quantum Information cannot be copied (no-cloning theorem).

\item{2)} Quantum Information cannot be deleted (no-erase theorem).

\item{3)} Heisenberg uncertainty principle.

\item{4)} Quantum superposition

\item{5)} Quantum entanglement

\item{6)} Quantum non-locality, EPR ``paradox''.

\end{itemize}

The corresponding logical requirements are:

\begin{itemize}

\item{1')} No contraction rule

\item{2')} No weakening rule

\item{3')} Non-distributivity, then no free contexts on both sides.

\item{4')} Connective \&=``superposition''

\item{5')} Connective @ =``entanglement''

\item{6')} The EPR rule

\end{itemize}

Requirements 1'--3' exclude all logics apart from Basic logic
\textbf{B} (and \textbf{BR}, \textbf{BL}, for more than two
qubits).

\textbf{B} satisfies the remaining requirements 4'--6'.

\vspace{1cm}

\noindent \textbf{Acknowledgements.} I wish to thank G.~Sambin for
useful discussions. I am grateful to the Organizers of DICE 2006,
where this work was presented for the first time, for their kind
interest, and encouragement, in particular  to Giuseppe Vitiello
for discussions and advices.


\bibliographystyle{unsrt}
\bibliography{liar_entanglement}

\end{document}